\documentclass[11pt]{cit_lab_mfr}

\usepackage[T1]{fontenc}    
\usepackage{url}            
\usepackage{booktabs}       
\usepackage{amsfonts}       
\usepackage{nicefrac}       
\usepackage{microtype}      
\usepackage{xspace}
\usepackage{array}
\usepackage{tabularx}
\usepackage{float}

\usepackage{amsmath, amssymb} 
\usepackage{graphics, graphicx} 
\usepackage{units} 
\usepackage{multicol}
\usepackage{mathtools}
\usepackage{tikz} 
\usetikzlibrary{shapes,decorations}
\usetikzlibrary{arrows.meta}
\usepackage{pgfplots} 
\pgfplotsset{compat=newest} 
\usepackage{comment}
\usepackage[normalem]{ulem} 
\usepackage[shortlabels]{enumitem}
\usepackage[linesnumbered,algoruled,lined]{algorithm2e}
\usepackage[overload]{empheq}
\usepackage{wrapfig}
\definecolor{mydarkblue}{rgb}{0,0.08,0.45}
\definecolor{confblue}{RGB}{45,85,135}
\definecolor{confbluebg}{RGB}{242,247,252}
\usepackage{fvextra}

\definecolor{siggreen}{RGB}{20,115,72}
\newcommand{\sig}[1]{\textcolor{siggreen}{$\mathbf{#1}$}}
\newcommand{\BlackBox}{\rule{1.5ex}{1.5ex}}  

\newcommand{\stkout}[1]{\ifmmode\text{\sout{\ensuremath{#1}}}\else\sout{#1}\fi}

\usepackage{contour}

\newcommand{\model}{PEAR\xspace}
\newcommand{\fullmodel}{\textbf{P}rogressive \textbf{E}vidence-Based \textbf{A}uto\textbf{R}esearch (PEAR)\xspace}
\newcommand{\strategyharness}{Evidence-driven AutoResearch\xspace}
\newcommand{\verifierladder}{Confidence-Gated Verifier Ladder\xspace}
\renewcommand{\bytedancelogopath}{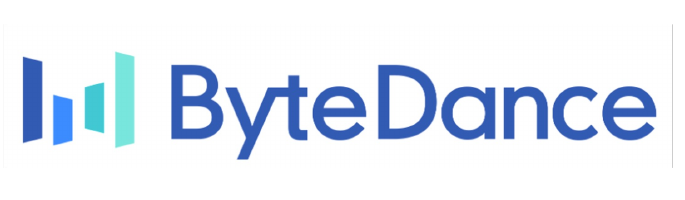}

\title{\model: Progressive Evidence-Based AutoResearch for Industrial Search Systems}
\author{Global E-Commerce Agentic Search Team}

\begin{document}
\abstract{
AutoResearch improves systems through iterative experimentation: agents propose candidate modifications, evaluate them, and use the results to guide subsequent exploration. Applying this paradigm to industrial search presents two challenges. (1) Common AutoResearch approaches follow a keep-if-better rule, retaining the highest-scoring candidate for subsequent experiments. Under non-stationary traffic, transient gains may be mistaken for persistent improvements, impairing reliable accumulation of search knowledge. (2) Candidate modifications can be evaluated at multiple fidelity levels, from low-cost proxies to online validation, differing in cost, objective alignment, and statistical reliability. Existing methods rely on individual signals or task-specific procedures, lacking a unified basis for using evidence across levels to guide search. We introduce \fullmodel with two complementary components. \strategyharness maintains an independent, hypothesis-guided research state for each strategy task within a predefined objective and intervention scope. Each state evolves through a Plan--Execute--Evaluate--Update transition that links experimentation to context-aware evidence interpretation and hypothesis revision. \verifierladder organizes evaluation into four levels of increasing fidelity: Offline Replay, Shadow-Traffic Evaluation, Rapid Online Evaluation, and Decision-Grade Online Evaluation. A unified confidence-based gate promotes candidates only when evidence supports a statistically significant positive effect, enabling broad low-cost exploration while reserving costly online experiments for promoted candidates. In a real-world industrial search system, strategies optimized with \model significantly increased Main Order/DAU by 2.7336\% and 3.2957\% relative to their respective baselines in two A/B experiments.
}

\maketitle


\section{Introduction}
Industrial search systems integrate query understanding, candidate retrieval, and result ranking to determine which results are presented to users. Recent work has substantially advanced individual components through instruction-aware retrieval and LLM-based reranking \citep{zhou2025contentrelevanceevaluatinginstruction,10.1145/3696410.3714620,10.1145/3696410.3714863,yoon2025acurankuncertaintyawareadaptivecomputation}. In production, however, these components operate within a multi-stage decision process in which models, ranking strategies, and business constraints interact. Shifts in user demand, content supply, and business objectives require these strategies to be continually revisited, while dependencies across stages make the effects of candidate modifications context-dependent and difficult to predict offline. Engineers therefore improve the system through iterative experimentation: they formulate hypotheses from observed system behavior, implement candidate strategies, evaluate them under real traffic, and use the results to guide subsequent iterations.

Recent advances in large language model (LLM) agents have demonstrated multi-step reasoning, tool use, and the ability to act in external environments \citep{yao2023react,schick2023toolformer,wang2024agentsurvey,yao2023treeofthoughts,park2023generativeagents,qian2024chatdev}, creating an opportunity to automate portions of this workflow. AutoResearch turns these capabilities into an iterative experimentation loop in which agents propose and implement candidate modifications, evaluate their outcomes, and use the resulting evidence to guide subsequent proposals \citep{huang2024mlagentbench,lu2024aiscientist,romeraparedes2024funsearch,boiko2023coscientist}. Recent work has applied this paradigm to search and recommendation systems, exploring interventions that range from ranking parameters to model architectures and reward functions \citep{wang2026self,cheng2026let}, and evaluating them using offline proxy metrics, model diagnostics, simulated user responses, and online A/B test outcomes \citep{kim2026self,lao2026agentx}. Collectively, these studies demonstrate the potential of agent-driven strategy optimization, while making experimental evidence the primary basis for candidate selection and subsequent exploration.

Reliably using evidence from both offline and online evaluations in AutoResearch for industrial search presents two challenges arising from non-stationary outcomes and heterogeneous evaluation procedures. \textbf{(1) How can an agent accumulate reliable experimental knowledge when candidate effects are context-dependent?} Many AutoResearch systems apply a keep-if-better rule, retaining the highest-scoring candidate as the reference for subsequent experiments \citep{romeraparedes2024funsearch,lu2024aiscientist}. This rule implicitly assumes that evaluation scores remain comparable across rounds. In industrial search, traffic composition and system state evolve over time; consequently, experiments conducted under different conditions may estimate candidate effects that are not directly comparable. A transient or context-specific gain may therefore be recorded as a persistent improvement, weakening hypothesis assessment and misdirecting subsequent exploration. \textbf{(2) How should evidence from different evaluation procedures govern candidate promotion?} Low-cost proxy evaluations support broad exploration, but their improvements may not translate into online gains. Online experiments provide evidence that is more directly aligned with deployment objectives, but sufficiently evaluating every candidate online is costly and risky \citep{li2018hyperband,kandasamy2017multifidelity,li2011unbiasedoffline,mann2019delayedproxies,kohavi2009controlled}. Existing approaches use individual evaluation signals or task-specific evaluation procedures \citep{wang2026self,kim2026self,cheng2026let,lao2026agentx}, leaving unclear how evidence from different procedures should jointly inform promotion decisions. The resulting challenge is to accumulate reliable experimental knowledge under changing contexts and make sound candidate-promotion decisions from heterogeneous evaluation evidence.

To address these challenges, we introduce \fullmodel, a framework for industrial search systems that couples persistent strategy states with progressive candidate evaluation through two complementary components.
\textbf{\strategyharness} maintains an independent research state for each strategy task, comprising the experimental context, current hypotheses, candidate intervention space, and accumulated evidence. The state evolves through a Plan--Execute--Evaluate--Update transition augmented with Bayesian-optimization-inspired reasoning. This reasoning balances refinement of evidence-supported regions with exploration of underexplored regions, while experimental results are interpreted within their contexts to support hypothesis revision and knowledge accumulation across experiments. 
\textbf{\verifierladder} organizes candidate evaluation into four progressively higher-fidelity levels through a unified confidence-based promotion gate: Offline Replay, Shadow-Traffic Evaluation, Rapid Online Evaluation, and Decision-Grade Online Evaluation. Evaluation results and promotion decisions from each level update the corresponding research state and inform subsequent search. Lower-cost evaluations support broad exploration, while costly online experiments are reserved for promoted strategies. In a real-world industrial search system, strategies optimized with \model significantly increased Main Order/DAU by 2.7336\% and 3.2957\% relative to their respective baselines in two A/B experiments at the Decision-Grade Online Evaluation stage.

\section{Contributions}

\begin{enumerate}
    \item We introduce \textbf{\strategyharness} to support reliable knowledge accumulation for industrial search optimization under non-stationary traffic. For each strategy task, it maintains an independent research state that links hypotheses and candidate interventions to their evaluation outcomes and experimental contexts. The state evolves through a Plan--Execute--Evaluate--Update transition, in which contextual evidence informs revisions to both the hypotheses and the candidate intervention space across experimental rounds.

    \item We develop \textbf{\verifierladder}, which organizes candidate evaluation into four levels of progressively greater fidelity and cost, from Offline Replay to Decision-Grade Online Evaluation. The ladder preserves each level's evaluation signal in a unified evidence record and applies a common confidence-based promotion rule, allowing low-cost evaluations to support broad exploration while reserving costly online experiments for statistically supported candidates. The resulting evidence and promotion decisions are returned to the research state to guide subsequent search.

    \item We deploy and evaluate \model in a real-world industrial search system using three strategy tasks. The evaluation documents evidence-driven hypothesis revision and, for two promoted candidates, directionally consistent positive effects from Shadow-Traffic Evaluation to Decision-Grade Online Evaluation. In two Decision-Grade Online A/B experiments, strategies optimized with \model significantly increased Main Order/DAU by 2.7336\% and 3.2957\% relative to their respective baselines.
\end{enumerate}

\section{Related Work}

\subsection{AutoResearch}

AutoResearch organizes candidate generation, experimentation, and evaluation into iterative research workflows \citep{tie2026autoresearch,huang2024mlagentbench,lu2024aiscientist}. A common update mechanism is the keep-if-better rule, which retains candidates that improve the observed experimental score as references for subsequent search \citep{romeraparedes2024funsearch,lu2024aiscientist}. The interpretation of these improvements depends on the comparability of evaluation outcomes across experiments.
Recent systems preserve experimental history to inform subsequent proposals. Meta-Harness optimizes harness code using the code, execution traces, and scores of prior candidates \citep{lee2026meta}. AiScientist maintains a persistent workspace that retains experimental artifacts across implementation, experimentation, and diagnosis \citep{chen2026toward}. AutoResearchClaw combines multi-agent deliberation, execution recovery, and cross-run experience accumulation to support iterative hypothesis generation and experimentation \citep{liu2026autoresearchclaw}. These approaches provide mechanisms for retaining and reusing information across experiments.

The keep-if-better rule uses observed candidate scores to guide subsequent search. In industrial search optimization, non-stationary traffic complicates the interpretation of these scores across experimental contexts. Reliable search knowledge accumulation requires distinguishing persistent improvements from transient gains over multiple experiments.

\subsection{AutoResearch in Search and Recommendation}

Recent studies apply AutoResearch to search and recommendation through task-specific candidate generation and evaluation procedures. Self-Evolving Recommendation System combines a fast offline loop for generating and filtering model candidates with a slower online loop for validating delayed business outcomes \citep{wang2026self}. Self-EvolveRec combines semantic feedback from a user simulator with model-diagnostic signals to guide candidate modifications across the recommendation pipeline \citep{kim2026self}. These studies use proxy evaluation and structured feedback to guide subsequent candidate generation.

Other systems incorporate online experimental feedback into iterative optimization. Sortify estimates the relationship between offline signals and online outcomes and maintains cross-round memory to inform subsequent parameter search \citep{cheng2026let}. AgentX connects candidate generation, implementation, online experimental evaluation, and experience feedback through specialized agents \citep{lao2026agentx}. These systems demonstrate how online outcomes can inform subsequent search in deployed search and recommendation systems.

Existing approaches organize candidate evaluation around individual signals or task-specific procedures. Proxy and online evaluation results differ in acquisition cost, objective alignment, and statistical reliability. A unified basis for using these results to guide candidate selection and subsequent search remains an open problem.


\section{Problem Setup}
\label{sec:problem-definition}
AutoResearch for industrial search optimization iteratively refines search strategies through candidate generation, experimentation, and evaluation.
Each strategy task $k$ specifies a predefined objective and intervention scope. A candidate strategy $c$ specifies a testable modification to the search system, such as a parameter configuration or a change to the ranking mechanism. Multiple strategy tasks may proceed concurrently within their respective intervention scopes.

At experimental round $t$, the agent generates a candidate set $\mathcal{A}_{k,t}$ for strategy task $k$. An evaluation procedure assigns scores to the candidates according to the optimization objective. The agent uses the resulting scores to refine subsequent proposals. Repeated candidate generation, evaluation, and proposal revision form an iterative optimization process.

In AutoResearch for industrial search systems, evaluations range from low-cost proxy evaluation to online validation. These evaluation sources differ in acquisition cost, objective alignment, and statistical reliability. Proxy evaluation supports broad candidate exploration at low cost. Online validation measures candidate effects through real-user outcomes and requires additional traffic and observation time. Let $\mathcal{S}(c)$ denote the set of evaluation stages applied to candidate $c$, and let $\mathcal{E}_{\ell}(c)$ denote the evidence acquired at stage $\ell$. The collected evidence is
\begin{equation}
    \mathcal{E}(c)
    =
    \left\{
        \mathcal{E}_{\ell}(c)
        \mid \ell\in\mathcal{S}(c)
    \right\}.
\end{equation}
For strategy task $k$, $\mathcal{E}_{k,t}^{(\ell)}$ denotes the set of stage-$\ell$ evidence records acquired in round $t$. Each result reflects its evaluation protocol and traffic context. Under non-stationary traffic, results from different rounds must be interpreted within their corresponding experimental contexts.

The objective is to optimize industrial search strategies through iterative experimentation guided by heterogeneous evaluation results. Evaluation results provide the basis for hypothesis revision, candidate selection, and subsequent search. This process requires reliable search knowledge accumulation across experimental rounds and sufficient online validation to support deployment decisions.

\section{Progressive Evidence-Based AutoResearch (PEAR)}

\begin{figure}
    \centering
    \includegraphics[width=0.95\linewidth]{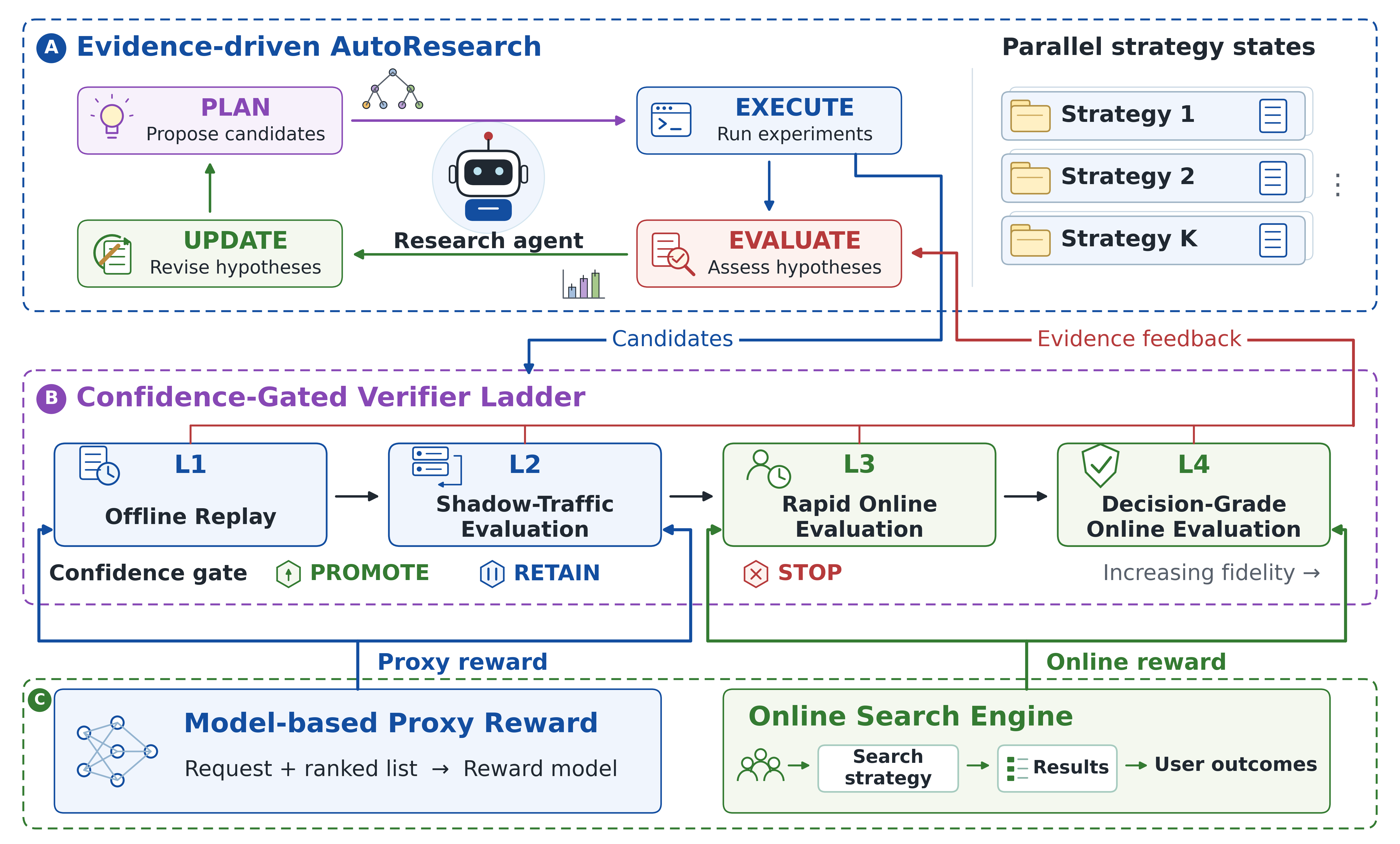}
    \caption{Overview of PEAR. \strategyharness maintains parallel strategy-level research states, while the \verifierladder progressively evaluates candidates, governs their promotion using evidence of increasing fidelity, and returns evaluation results to the corresponding research states.}
    \label{fig:model-overview}
\end{figure}

As illustrated in Figure~\ref{fig:model-overview}, \model couples persistent strategy states with progressive candidate evaluation through two complementary components: \strategyharness and \verifierladder.

\strategyharness maintains an independent, hypothesis-guided research state for each strategy task. Each state preserves the hypotheses, candidate history, and experimental contexts associated with that task. A Plan--Execute--Evaluate--Update transition connects candidate generation and experimentation with evidence interpretation and hypothesis revision.

\verifierladder organizes candidate evaluation into four progressively higher-fidelity levels: Offline Replay, Shadow-Traffic Evaluation, Rapid Online Evaluation, and Decision-Grade Online Evaluation. Unified evidence records and a confidence-based promotion gate connect these evaluation levels. Evaluation results and promotion decisions update the corresponding research states and inform subsequent search. Lower-cost evaluations support broad exploration, and costly online experiments are reserved for promoted candidates.

\subsection{\strategyharness}
\label{sec:harness}

\strategyharness associates optimization hypotheses with the evaluation evidence acquired for each strategy task. A hypothesis describes the expected effect of a candidate modification under specified conditions. Its associated evidence records the candidate, evaluation outcome, and experimental context. This association provides the basis for interpreting non-stationary evaluation results and revising hypotheses across experiments.

For strategy task $k$ at experimental round $t$, the research state is
\begin{equation}
\mathcal{H}_{k,t}
=
\left(
\mathcal{C}_{k},
\mathcal{M}_{k,t},
\Omega_{k,t},
\mathcal{D}_{k,t}
\right),
\label{eq:harness-state}
\end{equation}
where $\mathcal{C}_{k}$ specifies the search-system implementation, optimization objective, intervention scope, and experimental protocol. $\mathcal{M}_{k,t}$ contains the current hypotheses, $\Omega_{k,t}$ denotes the candidate intervention space within the specified scope, and $\mathcal{D}_{k,t}$ contains the accumulated evidence and its associations with the hypotheses. Independent states preserve the experimental context of each strategy task and support concurrent exploration across strategy tasks.

The research state evolves through a Plan--Execute--Evaluate--Update transition:
\begin{equation}
\mathcal{H}_{k,t}
\xrightarrow{\textsc{Plan}}
\mathcal{A}_{k,t}
\xrightarrow{\textsc{Execute}}
\Delta\mathcal{D}_{k,t}
\xrightarrow{\textsc{Evaluate}}
\mathcal{R}_{k,t}
\xrightarrow{\textsc{Update}}
\mathcal{H}_{k,t+1}.
\label{eq:autoresearch-loop}
\end{equation}
Here, $\mathcal{A}_{k,t}$ is the candidate set, $\Delta\mathcal{D}_{k,t}$ is the newly acquired evidence, and $\mathcal{R}_{k,t}$ records the evidence-based assessment of the current hypotheses.

\paragraph{\textsc{Plan}: Hypothesis-Guided Candidate Generation.}
The agent formulates testable hypotheses from the mechanisms of the industrial search system, the optimization objective, and accumulated experimental evidence. These hypotheses describe how candidate modifications are expected to affect the objective under the corresponding experimental conditions. Each candidate in $\mathcal{A}_{k,t}$ is associated with a hypothesis and an expected observation. Candidate generation includes interventions that examine individual effects, test interactions, or replicate previous observations.

Repeated refinement around previously promising candidates can concentrate exploration within a narrow region. To broaden candidate generation, the agent incorporates Bayesian-optimization-inspired reasoning:
\begin{equation}
\mathcal{A}_{k,t}
=
\mathcal{G}\!\left(
\mathcal{C}_{k},
\mathcal{M}_{k,t},
\mathcal{D}_{k,t},
\Omega_{k,t};
\mathcal{K}_{\mathrm{BO}}
\right),
\qquad
\mathcal{A}_{k,t}\subseteq\Omega_{k,t},
\label{eq:knowledge-augmented-proposal}
\end{equation}
where $\mathcal{G}$ denotes the agent's candidate-generation procedure and $\mathcal{K}_{\mathrm{BO}}$ denotes Bayesian-optimization-inspired reasoning principles. Conditioned on the search-system context, current hypotheses, and accumulated evidence, this procedure combines refinement of evidence-supported regions with exploration of underexplored regions. These principles guide qualitative reasoning about observed effects and uncertainty rather than specifying a fitted numerical optimization model.

\paragraph{\textsc{Execute}: Experiment Execution and Evidence Collection.}
The agent invokes stage-specific experimentation tools to apply candidate modifications within the corresponding evaluation environments and initiate experiments under \verifierladder. As candidates progress through the evaluation stages, the agent retrieves the evaluation results and promotion decisions produced by the corresponding verifiers. These outputs form the newly acquired evidence $\Delta\mathcal{D}_{k,t}$, with each record linked to its candidate, hypothesis, and experimental context.

\paragraph{\textsc{Evaluate}: Contextual Hypothesis Assessment.}
The agent assesses each hypothesis using its associated evidence in $\mathcal{D}_{k,t}\cup\Delta\mathcal{D}_{k,t}$. This assessment considers the candidate modifications, evaluation stages, and experimental contexts underlying the observed effects. Consistent observations provide support for a hypothesis. Conflicting observations motivate examination of its assumptions and applicability conditions. Inconclusive observations remain unresolved evidence for further experimentation. The assessment record $\mathcal{R}_{k,t}$ preserves these distinctions for hypothesis revision.

\paragraph{\textsc{Update}: Evidence-Based Hypothesis Revision.}
The assessment record updates the hypotheses and the candidate intervention space:
\begin{equation}
\begin{aligned}
(\mathcal{M}_{k,t+1},\Omega_{k,t+1})
&=
\mathcal{U}
(\mathcal{M}_{k,t},\Omega_{k,t},\mathcal{R}_{k,t}),\\
\mathcal{D}_{k,t+1}
&=
\mathcal{D}_{k,t}\cup\Delta\mathcal{D}_{k,t},
\end{aligned}
\label{eq:harness-update}
\end{equation}
where $\mathcal{U}$ denotes evidence-based revision of the hypotheses and candidate intervention space. Supported hypotheses guide further refinement and replication. Conflicting evidence motivates revised hypotheses and alternative candidate modifications within the specified intervention scope. Evaluation results and promotion decisions remain available in the updated state, allowing subsequent proposals to build on accumulated evidence and its experimental context.

\subsection{\verifierladder}
\label{sec:verifier-ladder}

\verifierladder organizes heterogeneous candidate evaluations into four levels: Offline Replay (L1), Shadow-Traffic Evaluation (L2), Rapid Online Evaluation (L3), and Decision-Grade Online Evaluation (L4). \textbf{All four levels are aligned with the same optimization objective but differ in signal source, evaluation fidelity, acquisition cost, traffic exposure, and observation horizon.} L1 and L2 use a model-based proxy score that estimates this objective from ranking outputs, and L3 and L4 use the objective directly observed from online user outcomes. Each strategy task instantiates an ordered sequence of active levels according to data availability and validation requirements. A unified confidence-based promotion rule connects consecutive levels in this task-specific sequence. Each active level produces evaluation results under its corresponding experimental protocol. Evaluation results and promotion decisions provide evidence for hypothesis assessment and subsequent search.

\subsubsection{Unified Promotion Policy}

The promotion policy uses the estimated candidate effect and its confidence interval at each evaluation stage. For candidate $c$ at stage $\ell$, the verifier produces
\begin{equation}
\mathcal{E}_{\ell}(c)
=
\big(
c,b_{\ell},m_{\ell},
n_{\ell}^{c},n_{\ell}^{b_{\ell}},
\widehat{\Delta}_{\ell}(c),
\mathrm{CI}_{\ell}(c),
v_{\ell}(c)
\big),
\end{equation}
where $\mathcal{E}_{\ell}(c)$ is the evidence record produced for candidate $c$ at stage $\ell$, which packages the comparison and its statistical summary into a single object. Within it, $b_{\ell}$ is the stage-specific baseline and $m_{\ell}$ is the evaluation signal. The quantities $n_{\ell}^{c}$ and $n_{\ell}^{b_{\ell}}$ denote the numbers of valid evaluation units for the candidate and baseline. $\widehat{\Delta}_{\ell}(c)$ denotes the estimated candidate effect, $\mathrm{CI}_{\ell}(c)$ its confidence interval, and $v_{\ell}(c)$ the promotion decision. L1 and L2 use model-based proxy scores. L3 and L4 use observed online outcomes. This record is the unit of evidence accumulated in the research state and interpreted together with its experimental context.

Let $\widehat{\mu}_{\ell}(c)$ and $\widehat{\mu}_{\ell}(b_{\ell})$ denote the estimated values of the evaluation signal for the candidate and baseline. With the signal oriented so that larger values indicate better outcomes, the estimated effect is
\begin{equation}
\widehat{\Delta}_{\ell}(c)
=
\widehat{\mu}_{\ell}(c)
-
\widehat{\mu}_{\ell}(b_{\ell}).
\end{equation}
The confidence interval is centered on this effect and widened by its standard error,
\begin{equation}
\mathrm{CI}_{\ell}(c)
=
\Bigl[\,
\widehat{\Delta}_{\ell}(c)-t_{\nu_{\ell},\,1-\alpha/2}\,\mathrm{SE}_{\ell}(c),
\;\;
\widehat{\Delta}_{\ell}(c)+t_{\nu_{\ell},\,1-\alpha/2}\,\mathrm{SE}_{\ell}(c)
\,\Bigr],
\label{eq:confidence-interval}
\end{equation}
where $\mathrm{SE}_{\ell}(c)$ is the standard error of the effect, $t_{\nu_{\ell},\,1-\alpha/2}$ is the critical value at significance level $\alpha$, and $\nu_{\ell}$ the associated degrees of freedom. The evidence representation and promotion rule are shared across levels, whereas effect estimation follows the sampling structure of each evaluation protocol. The shared representation retains stage-specific signal definitions and confidence intervals rather than combining heterogeneous results into a single score. The stage-specific effect estimators and confidence computations are detailed in Appendix~\ref{app:confidence}.

Let $\mathrm{CI}_{\ell}(c)=[L_{\ell}(c),U_{\ell}(c)]$ denote the confidence interval at the prescribed confidence level. The promotion decision is
\begin{equation}
v_{\ell}(c)
=
\begin{cases}
\mathsf{PROMOTE}, & L_{\ell}(c)>0,\\
\mathsf{RETAIN}, & L_{\ell}(c)\leq 0\leq U_{\ell}(c),\\
\mathsf{STOP}, & U_{\ell}(c)<0.
\end{cases}
\end{equation}
$\mathsf{PROMOTE}$ indicates statistically significant evidence of a positive effect and advances the candidate to the next evaluation stage. At L4, it indicates that the candidate satisfies the promotion rule for deployment consideration. $\mathsf{RETAIN}$ indicates inconclusive evidence and retains the candidate for further evaluation at the current stage. $\mathsf{STOP}$ indicates statistically significant evidence of a negative effect and terminates the candidate's progression through the ladder.

The promotion decision progressively narrows the candidate set along the active evaluation sequence. For strategy task $k$, let $\mathcal{L}_{k}=(\ell_{k,1},\ldots,\ell_{k,J_k})$ denote its ordered sequence of active levels, where $\ell_{k,j}\in\{1,2,3,4\}$ and $\ell_{k,j}<\ell_{k,j+1}$. The candidates entering the first active level are those generated by \strategyharness, $\mathcal{A}_{k,t}^{[1]}=\mathcal{A}_{k,t}$, and each active level advances only the candidates it promotes:
\begin{equation}
\mathcal{A}_{k,t}^{[j+1]}
=
\bigl\{\,
c\in\mathcal{A}_{k,t}^{[j]}
\;:\;
v_{\ell_{k,j}}(c)=\mathsf{PROMOTE}
\,\bigr\},
\qquad
j\in\{1,\ldots,J_k\},
\label{eq:candidate-filtering}
\end{equation}
which yields the nested selection $\mathcal{A}_{k,t}^{[1]}\supseteq\mathcal{A}_{k,t}^{[2]}\supseteq\cdots\supseteq\mathcal{A}_{k,t}^{[J_k+1]}$. A candidate reaches an active level only after being promoted at every preceding level in the task-specific sequence. For deployment consideration, the sequence terminates at Decision-Grade Online Evaluation (L4), and $\mathcal{A}_{k,t}^{[J_k+1]}$ contains the candidates promoted at that level. Thus, lower-cost active levels screen candidates before costly online evaluation.

Evaluation results and promotion decisions are returned to the corresponding research state together with their experimental contexts. For strategy task $k$ in experimental round $t$, $\mathcal{E}_{k,t}^{(\ell)}$ collects the records of candidates evaluated at stage $\ell$. These records support evidence interpretation across experiments while preserving the conditions under which each result was obtained.

\subsubsection{Model-Based Proxy Evaluation}
\label{sec:reward-model}

L1 and L2 evaluate candidate ranking outputs without exposing users to those outputs. A list-level evaluation model provides a proxy score for each candidate. Given request context $x$ and the ranked list $L^{g}(x)$ produced by strategy $g$, the proxy evaluation signal is
\begin{equation}
m_{\ell}(g;x)=R_{\psi}\!\left(x,L^{g}(x)\right),
\qquad \ell\in\{1,2\},
\end{equation}
where $R_{\psi}$ estimates the online objective from the request context and the complete ranked list. The model accounts for the composition and relative positions of multiple content formats. It therefore evaluates the combined ranking output rather than individual content scores in isolation.

The resulting proxy scores provide the evaluation signal for estimating candidate effects at L1 and L2. Both stages use the same scoring model but evaluate candidates under different request and execution conditions. Offline Replay uses fixed historical requests. Shadow-Traffic Evaluation uses duplicated real-time requests in the online execution environment.

\subsubsection{L1: Offline Replay}

L1 evaluates candidate modifications on a fixed set of historical requests. The candidate and baseline ranking procedures process the recorded request contexts and produce their respective ranked lists. The evaluation model assigns a proxy score to each list. For strategy $g\in\{c,b_1\}$, the estimated proxy score is
\begin{equation}
\widehat{\mu}_{1}(g)=\widehat{\mathbb{E}}_{x\sim\mathcal{X}_{\mathrm{hist}}}\!\left[m_{1}(g;x)\right],
\label{eq:offline-replay-effect}
\end{equation}
where $\mathcal{X}_{\mathrm{hist}}$ is the historical request set over which the proxy signal is averaged. Because the candidate and baseline process the same historical requests, L1 estimates the candidate effect from request-level paired differences.

The verifier forms the effect $\widehat{\Delta}_{1}(c)$ and its confidence interval $\mathrm{CI}_{1}(c)$ from these scores under the Offline Replay protocol. This evidence supports low-cost candidate screening. The promoted subset $\mathcal{A}_{k,t}^{(2)}$ proceeds to Shadow-Traffic Evaluation.

\subsubsection{L2: Shadow-Traffic Evaluation}

L2 evaluates candidate modifications using duplicated real-time requests in the industrial search system. Candidate ranking procedures execute with current request contexts and online features. Their outputs are used for evaluation without replacing the results shown to users. The per-request proxy signal $m_{2}(g;x)$ follows the same model-based definition, with $L^{g}(x)$ produced in the online execution environment. The scoring model is shared with L1, but its inputs reflect current requests, online features, and the ranking outputs generated under these conditions. For strategy $g\in\{c,b_2\}$, the estimated proxy score is
\begin{equation}
\widehat{\mu}_{2}(g)=\widehat{\mathbb{E}}_{x\sim\mathcal{X}_{\mathrm{shadow}}^{g}}\!\left[m_{2}(g;x)\right],
\label{eq:shadow-traffic-effect}
\end{equation}
where $\mathcal{X}_{\mathrm{shadow}}^{g}$ is the set of requests received by the shadow-traffic group for strategy $g$. Candidate and baseline statistics are computed from independent shadow-traffic groups that receive requests replicated from the live traffic stream, without request-level pairing across groups.

The verifier forms the effect $\widehat{\Delta}_{2}(c)$ and its confidence interval $\mathrm{CI}_{2}(c)$ from these scores, which determine candidate promotion under the unified policy. This stage examines candidate performance under current operating conditions while retaining model-based proxy evaluation. The promoted subset $\mathcal{A}_{k,t}^{(3)}$ proceeds to Rapid Online Evaluation.

\subsubsection{L3: Rapid Online Evaluation}

L3 applies candidate modifications to limited randomized traffic. Users in the candidate group receive results produced by the candidate strategy, and users in the baseline group receive results produced by the baseline strategy. The experiment collects observed user outcomes over a short observation window. These outcomes replace model-based proxy scores as the evaluation signal.

Let $\mathcal{N}_{3}^{g}$ denote the evaluation units observed under strategy $g$, and let $y_{3}(g;u)$ denote the observed objective signal for unit $u$. For strategy $g\in\{c,b_3\}$, the estimated value is
\begin{equation}
\widehat{\mu}_{3}(g)=\widehat{\mathbb{E}}_{u\sim\mathcal{N}_{3}^{g}}\!\left[y_{3}(g;u)\right],
\label{eq:rapid-online-effect}
\end{equation}
where the expectation is taken over the units randomly assigned to each strategy. Evaluation units and outcome definitions follow the online experimental protocol. The verifier forms the effect $\widehat{\Delta}_{3}(c)$ and its confidence interval $\mathrm{CI}_{3}(c)$ under that protocol and applies the unified promotion policy. This stage assesses whether candidates selected through proxy evaluation produce positive effects in real-user outcomes. The promoted subset $\mathcal{A}_{k,t}^{(4)}$ proceeds to Decision-Grade Online Evaluation.

\subsubsection{L4: Decision-Grade Online Evaluation}

L4 evaluates promoted candidates through online experiments with broader randomized traffic and a longer observation window. Candidate and baseline groups provide observed outcomes aligned with the optimization objective. The expanded experiment covers a wider range of traffic conditions and temporal variation.

Using the same notation for observed objective signals, the estimated value is
\begin{equation}
\widehat{\mu}_{4}(g)=\widehat{\mathbb{E}}_{u\sim\mathcal{N}_{4}^{g}}\!\left[y_{4}(g;u)\right],
\qquad g\in\{c,b_4\},
\label{eq:decision-grade-effect}
\end{equation}
where $\mathcal{N}_{4}^{g}$ contains the evaluation units used in Decision-Grade Online Evaluation and $y_{4}(g;u)$ is the objective signal measured over its observation window. The resulting effect $\widehat{\Delta}_{4}(c)$ and its confidence interval $\mathrm{CI}_{4}(c)$ support the final deployment decision. Candidates satisfying the promotion rule at L4 are eligible for deployment consideration. Evaluation results and final decisions update the corresponding research states and inform subsequent search.

\section{Experiments}

\subsection{Experimental Setup and Research Questions}

The experiments evaluate \model in a real-world industrial search system. Three anonymized strategy tasks provide records of candidate exploration and evaluation. These records include experimental configurations, evaluation results, and hypothesis revisions. A/B results from Decision-Grade Online Evaluation are reported for strategies identified for two of these tasks.

The evaluation addresses three research questions:

\noindent\textbf{RQ1: Evidence-driven hypothesis revision.}
How are accumulated evaluation results incorporated into hypothesis revision and subsequent candidate generation?\par

\noindent\textbf{RQ2: Progressive candidate evaluation.}
How do proxy evaluation results relate to observed online outcomes, and how do selected candidates behave across evaluation stages?\par

\noindent\textbf{RQ3: Online effectiveness.}
Do strategies optimized with \model improve the online optimization objective during Decision-Grade Online Evaluation?
\paragraph{Implementation Detail.}
Experiments at L1--L3 run on timescales of hours, and L4 experiments run on timescales of days. Main Order/DAU is the primary outcome reported for the final online A/B experiments; the remaining reported outcomes are auxiliary business metrics for the e-commerce search setting. Traffic conditions, observation windows, and evaluation signals differ across stages. Cross-stage analysis therefore examines the direction of candidate effects rather than directly comparing their magnitudes.
For industrial search optimization, each strategy task instantiates an ordered sequence of active evaluation levels according to data availability and validation requirements. Candidates must satisfy the promotion criterion at each active level before advancing to the next level in that sequence. In the following experiments, data-access restrictions across regions preclude the use of Offline Replay (L1), so the evaluated sequence begins at Shadow-Traffic Evaluation (L2) and the subsequent experimental sections do not involve L1 validation. Although L1 is not used in these experiments, it has assisted algorithm engineers in optimizing other strategies in production.

\subsection{RQ1: Evidence-Driven Hypothesis Revision}

\paragraph{Candidate exploration and selection.}

\begin{table}[htbp]
    \centering
    \caption{L2 search scale and candidate selection. Unique configurations are deduplicated within each phase; promotion rate is the number of candidates promoted from L2 divided by the number of experimental groups.}
    \label{tab:search-scale}
    \small
    \renewcommand{\arraystretch}{1.08}
    \begin{tabular*}{0.92\linewidth}{@{\extracolsep{\fill}}lrrrrr@{}}
        \toprule
        Task phase & Experimental rounds & Groups & Unique configs. & Promoted from L2 & Promotion rate \\
        \midrule
        A & 19 & 133 & 20 & 2 & 1.50\% \\
        B--I & 12 & 84 & 23 & 5 & 5.95\% \\
        B--II & 2 & 14 & 6 & 1 & 7.14\% \\
        B--III & 21 & 147 & 53 & 2 & 1.36\% \\
        C & 16 & 76 & 54 & 3 & 3.95\% \\
        \bottomrule
    \end{tabular*}
\end{table}

Table~\ref{tab:search-scale} summarizes candidate exploration at L2 for three anonymized strategy tasks. Each task evaluates multiple configurations and retains a small subset for online evaluation. Task B comprises three consecutive search phases: the first two explore local parameter neighborhoods, and Phase~III expands the candidate range based on accumulated evaluation results. The following trajectories examine how this evidence informs hypothesis revision and subsequent candidate generation.

\paragraph{Hypothesis revision through controlled interventions.}
\begin{table}[htbp]
    \centering
    \caption{Representative hypothesis-revision trajectory for Task A. The table shows four early experimental rounds from a 19-round search; $G=(w,q)$ denotes the global window and capacity constraint.}
    \label{tab:mechanism-trajectory}
    \footnotesize
    \setlength{\tabcolsep}{4pt}
    \renewcommand{\arraystretch}{1.14}
    \begin{tabularx}{\linewidth}{>{\raggedright\arraybackslash}p{0.27\linewidth}>{\raggedright\arraybackslash}p{0.32\linewidth}>{\raggedright\arraybackslash}X}
        \toprule
        Experimental round / research question & Controlled interventions & Evidence-driven update \\
        \midrule
        \textbf{1.} Which constraint axis accounts for the observed gain? & Apply one-axis changes to global and format-specific constraints; also test the coupled global change $G=(3,2)$. & Only $G=(3,2)$ yields a significant positive order signal. The effects of increasing $w$, increasing $q$, and experimental variation remain entangled. \\
        \addlinespace[2pt]
        \textbf{2.} Can $G=(3,2)$ be reproduced, and which variable explains its gain? & Replicate $G=(3,2)$; compare window-only $G=(3,1)$, capacity-only $G=(2,2)$, and boundary settings including $G=(3,3)$. & $G=(3,3)$ produces a stronger signal. Replications of $G=(3,2)$ yield mixed directional effects and indicate potential degradation in the exit metric. The next experimental round prioritizes capacity and replication. \\
        \addlinespace[2pt]
        \textbf{3.} Is the gain driven by the coupled change or by capacity alone? & Replicate $G=(3,3)$ and $G=(3,2)$; test capacity-only $G=(2,2)$ and wider-window boundary settings. & $G=(2,2)$ achieves the strongest observed result. The revised hypothesis attributes the gain primarily to capacity relaxation without requiring a wider window. \\
        \addlinespace[2pt]
        \textbf{4.} Are the marginal effects of window and capacity stable? & Concentrate replications on $G=(2,2)$, $G=(3,2)$, and $G=(3,3)$; retain window-only $G=(3,1)$ as a counterfactual. & $G=(3,3)$ remains directionally favorable on order and exit signals but is not significant. The capacity-centered hypothesis is retained for further replication. \\
        \bottomrule
    \end{tabularx}
\end{table}

Task A illustrates how controlled interventions distinguish competing explanations for an observed gain. The task optimizes constraints on e-commerce result density. Let $G=(w,q)$ denote the global constraint, where $w$ is the window size and $q$ is the maximum content capacity within that window; the deployed baseline is $G=(2,1)$. Table~\ref{tab:mechanism-trajectory} traces four representative early experimental rounds. Other constraint families remain at their baseline values after initial screening.

The initial gain from $G=(3,2)$ leaves the effects of window size and capacity unresolved. The next experimental round separates these factors and replicates the coupled setting. Mixed directional effects across replications motivate further tests of capacity relaxation. In Experimental Round~3, the capacity-only setting $G=(2,2)$ achieves the strongest observed result, shifting the hypothesis toward capacity relaxation rather than window expansion. Experimental Round~4 then examines the consistency of this hypothesis through additional controlled comparisons and replication.

\paragraph{Hypothesis revision from online evaluation.}
Task B optimizes position-dependent exposure settings for e-commerce search results. Table~\ref{tab:cross-round} summarizes three search phases and their outcomes from Decision-Grade Online Evaluation.

\begin{table}[htbp]
    \centering
    \caption{Outcomes from Decision-Grade Online Evaluation across search phases for Task B.}
    \label{tab:cross-round}
    \small
    \renewcommand{\arraystretch}{1.08}
    \begin{tabular*}{0.92\linewidth}{@{\extracolsep{\fill}}lrrrrl@{}}
        \toprule
        Phase & Experimental rounds & Groups & Unique configs. & Promoted from L2 & L4 outcome \\
        \midrule
        I & 12 & 84 & 23 & 5 & No significant gain \\
        II & 2 & 14 & 6 & 1 & Positive, not significant \\
        III & 21 & 147 & 53 & 2 & Significant order gain \\
        \bottomrule
    \end{tabular*}
\end{table}

\begin{figure}[htbp]
    \centering
    \includegraphics[width=0.8\linewidth]{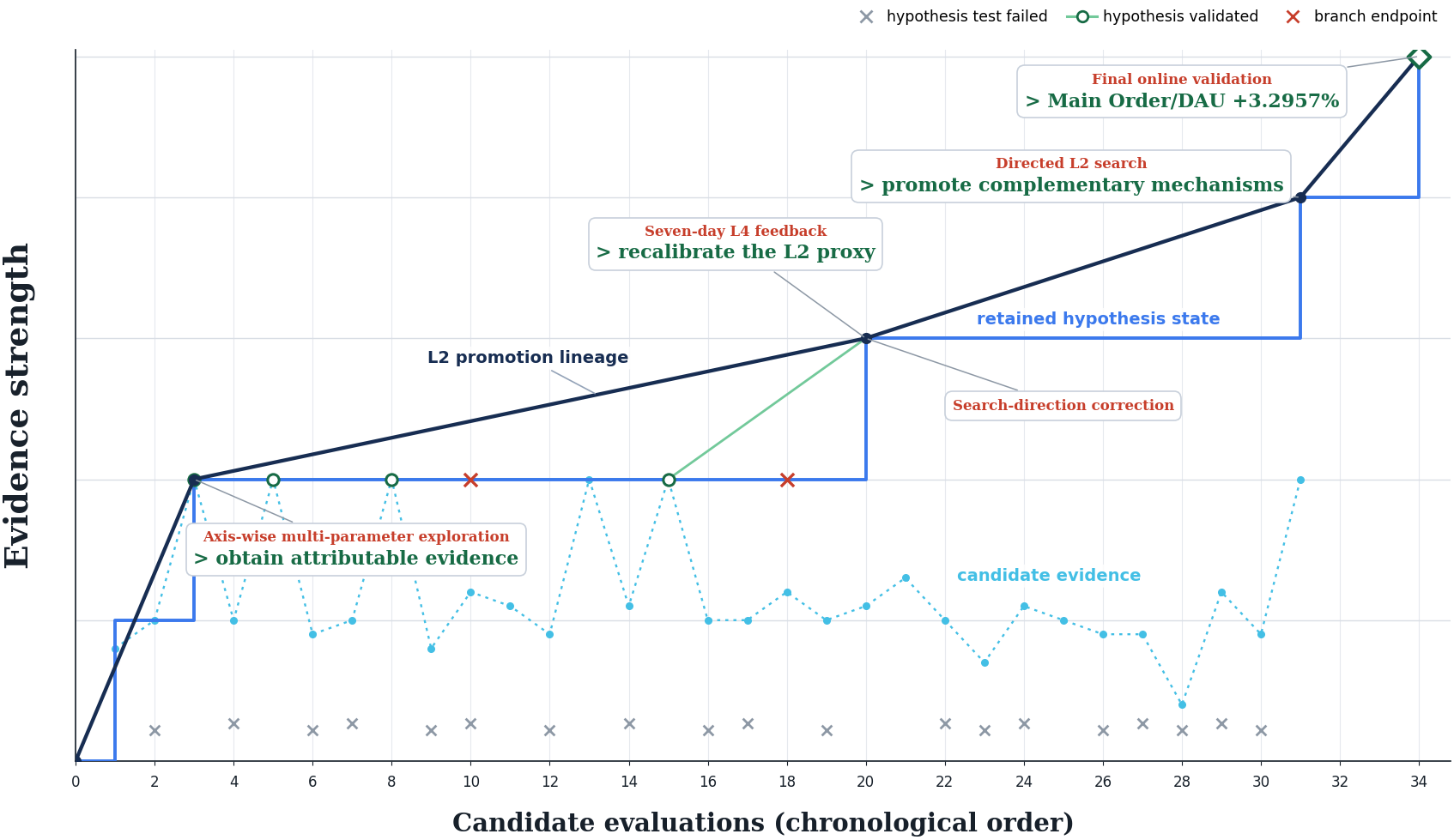}
    \caption{Hypothesis-driven search trajectory for Task B. Candidate evaluations update a retained hypothesis state through validation, rejection, and branch termination until Decision-Grade Online Evaluation.}
    \label{fig:hypothesis-driven-search}
\end{figure}

In the first two phases, candidates promoted from L2 do not achieve significant improvements at L4. These online evaluation results are incorporated into the research state to reassess the hypotheses underlying candidate generation. The revised hypotheses guide further exploration at L2, expanding the candidate range beyond the initial parameter neighborhoods. Figure~\ref{fig:hypothesis-driven-search} illustrates this exploration process: the agent uses accumulated evaluation evidence to revise hypotheses, terminate unsupported branches, and redirect candidate generation. In Phase~III, the revised exploration identifies candidates for further online evaluation, one of which subsequently achieves a significant improvement at L4. The trajectory illustrates how evidence from Decision-Grade Online Evaluation is incorporated into hypothesis revision and subsequent candidate generation, connecting candidate promotion with continued search.

\subsection{RQ2: Progressive Candidate Evaluation}
\paragraph{Aggregate agreement between proxy predictions and online outcomes.}
An independent controlled experiment compares mean predicted and observed click, order, and GMV outcomes at a common session granularity. Table~\ref{tab:proxy-agreement} reports the signed relative errors between these aggregate means. The order component has the smallest absolute relative error, with a signed value of $-3.49\%$. This comparison assesses aggregate agreement rather than the accuracy of predicted effects for individual candidate interventions.

\begin{table}[htbp]
    \centering
    \caption{Signed relative errors between the aggregate means of proxy predictions and observed online outcomes.}
    \label{tab:proxy-agreement}
    \small
    \renewcommand{\arraystretch}{1.10}
    \begin{tabular}{lccc}
        \toprule
        & Click & Order & GMV \\
        \midrule
        Signed relative error & $-16.02\%$ & $\mathbf{-3.49\%}$ & $-18.23\%$ \\
        \bottomrule
    \end{tabular}
\end{table}

\paragraph{Candidate effects across evaluation stages.}
Table~\ref{tab:stage-transfer} follows two Task A candidates from L2 through L3 to L4. Both traced candidates have positive order proxy effects at L2 and retain positive observed effects at L3 and L4. These cases illustrate directional consistency from proxy-based screening to Decision-Grade Online Evaluation.

\begin{table}[htbp]
    \centering
    \caption{Candidate effects across evaluation stages for two Task A candidates. L2 reports changes in the order proxy, L3 reports changes in observed orders, and L4 reports changes in Main Order/DAU.}
    \label{tab:stage-transfer}
    \small
    \renewcommand{\arraystretch}{1.12}
    \begin{tabularx}{\linewidth}{l>{\raggedright\arraybackslash}X>{\raggedright\arraybackslash}X>{\raggedright\arraybackslash}X}
        \toprule
        Candidate & L2: Shadow-Traffic Evaluation & L3: Rapid Online Evaluation & L4: Decision-Grade Online Evaluation \\
        \midrule
        A & Order proxy $+45.88\%$ & Observed order $+15.48\%$ & Main Order/DAU $+2.73\%$ \\
        B & Order proxy $+61.64\%$ & Observed order $+29.76\%$ & Main Order/DAU $+2.16\%$ \\
        \bottomrule
    \end{tabularx}
\end{table}

Together, the aggregate comparison and the two candidate trajectories provide illustrative evidence that proxy evaluation can support candidate screening, while online evaluation remains necessary to establish effects under real-user traffic.

\subsection{RQ3: Online Effectiveness}

Table~\ref{tab:online-ab} reports Decision-Grade Online Evaluation results for strategies optimized with \model for two search optimization tasks. Relative to their respective baselines, the strategies for Task A and Task B significantly improve Main Order/DAU by 2.7336\% and 3.2957\%, respectively. Both strategies also achieve significant gains in ASN, SKU Order/DAU, Main OPMS, and SKU OPMS.

\begin{table}[htbp]
    \centering
    \caption{Relative metric changes in online A/B evaluations of strategies optimized with \model for two search optimization tasks. All values are relative to the corresponding baseline; dark-green boldface indicates statistical significance at $p<0.05$.}
    \label{tab:online-ab}
    \small
    \setlength{\tabcolsep}{4.5pt}
    \renewcommand{\arraystretch}{1.18}
    \resizebox{\linewidth}{!}{%
    \begin{tabular}{lrrrrrrrr}
        \toprule
        Task & SearchPV/DAU & ASN & GMV/DAU & Main Order/DAU & SKU Order/DAU & PayPV/PV & Main OPMS & SKU OPMS \\
        \midrule
        A & $+0.0063\%$ & \sig{+1.1512\%} & $+2.2136\%$ & \sig{+2.7336\%} & \sig{+2.7259\%} & $+1.0441\%$ & \sig{+2.7407\%} & \sig{+2.7375\%} \\
        B & $+0.0470\%$ & \sig{+1.3841\%} & $+1.6633\%$ & \sig{+3.2957\%} & \sig{+2.9826\%} & \sig{+1.6298\%} & \sig{+3.2595\%} & \sig{+2.9448\%} \\
        \bottomrule
    \end{tabular}
    }
\end{table}

\section{Conclusion}

\model is an AutoResearch framework for industrial search optimization under non-stationary outcomes and heterogeneous evaluation signals. \strategyharness associates hypotheses with evidence and its experimental context to support reliable search knowledge accumulation. \verifierladder connects four evaluation levels through a unified confidence-based promotion gate. The two components jointly turn evaluation results into signals that guide both candidate promotion and hypothesis revision, linking lower-cost exploration with Decision-Grade Online Evaluation.

Experiments in a real-world industrial search system show how accumulated evidence informs controlled interventions and how online evaluation results redirect subsequent search. Strategies optimized with \model increased Main Order/DAU by 2.7336\% and 3.2957\% relative to their respective baselines in two A/B experiments at the Decision-Grade Online Evaluation stage. These results support the use of contextual evidence and progressive evaluation for iterative industrial search optimization. Future work will examine broader intervention scopes, including model architectures and training procedures.

\section{Limitations}
The current work is confined to a single e-commerce general search environment, and \model's generality across other search platforms, geographic regions, and evaluation configurations remains to be established. Where compliance constraints preclude Offline Replay (L1), evaluation proceeds from L2 onward while retaining the higher-fidelity validation stages. The search space is currently restricted to strategy-level parameters; its extension to model parameters, architectures, and training procedures is left to future work. \verifierladder also depends on the compute clusters, deployed evaluation and search models, and live traffic of the underlying search environment, which constrains its applicability in settings where such resources are unavailable.

\section{Ethical Considerations}
In the evaluated deployment, AutoResearch is employed to optimize an e-commerce general search system. As optimization feedback, the agents receive model-based proxy scores produced by the list-level evaluation model and aggregate effect estimates of the objective signal from online experiments, rather than user-level records, personal identifiers, sensitive attributes, or other sensitive information. All data used for evaluation and experimentation are processed under applicable data-governance and compliance requirements. The optimization loop therefore operates on compliant evaluation signals within the existing authorized experimentation pipeline and does not require additional access to personal or sensitive data.

Before any candidate is exposed to real-user traffic, the proposed change and supporting evidence undergo human review and require explicit human confirmation; automated promotion within the Verifier Ladder cannot independently initiate such exposure.

\section{Acknowledgments}
We thank our colleagues on the E-commerce General Search, Architecture, Product, and Data Science teams for their invaluable contributions to the design, development, and deployment of the systems described in this work. We are grateful to the AutoSearch tooling team for building the compute cluster and infrastructure that enabled large-scale AutoResearch for industrial search system optimization.

\section*{Author List}

\noindent\textbf{Global E-Commerce Agentic Search Team:} Yifan Wang\textsuperscript{*}, Shipeng Zhu, Fei Xiong, Yuqin Yang, Yonghui Huang, Kunyao Wu, Yue Wang, Weichao Meng, Yu Gong\textsuperscript{\textdagger}.

\noindent\textsuperscript{*}: First author, \email{yifanwang993w@gmail.com}\\
\textsuperscript{\textdagger}: Corresponding author, \email{gy910210@gmail.com}

\bibliographystyle{apalike}
\bibliography{references.bib}

\appendix

\section{Confidence Computation for the Unified Promotion Policy}
\label{app:confidence}

This appendix specifies the stage-specific estimators used by the unified promotion policy (Section~\ref{sec:verifier-ladder}) to obtain the candidate effect $\widehat{\Delta}_{\ell}(c)$ and its confidence interval $\mathrm{CI}_{\ell}(c)$. The evidence schema and promotion rule are shared across levels, while the estimator follows the sampling structure of each evaluation protocol. L1 uses request-aligned paired observations. L2 uses independent shadow-traffic groups, and L3 and L4 use independent randomized online groups.

\paragraph{Paired estimation for Offline Replay (L1).}
Let $x_i$ denote the $i$-th historical request shared by candidate $c$ and baseline $b_1$, and define the request-level proxy-score difference as
\begin{equation}
d_i=m_1(c;x_i)-m_1(b_1;x_i),
\qquad i\in\{1,\ldots,n\}.
\end{equation}
The L1 effect, variance of the paired differences, standard error, and degrees of freedom are
\begin{equation}
\widehat{\Delta}_1(c)=\bar d=\frac{1}{n}\sum_{i=1}^{n}d_i,
\qquad
s_d^2=\frac{1}{n-1}\sum_{i=1}^{n}(d_i-\bar d)^2,
\end{equation}
\begin{equation}
\mathrm{SE}_1(c)=\frac{s_d}{\sqrt{n}},
\qquad
\nu_1=n-1.
\end{equation}
This paired estimator controls for request-level variation because both strategies are evaluated on the same historical requests.

\paragraph{Independent-group estimation for L2--L4.}
For $\ell\in\{2,3,4\}$ and strategy $g\in\{c,b_{\ell}\}$, let $x_{g,i}^{(\ell)}$ denote its $i$-th valid evaluation unit and let $n_{g,\ell}$ be the number of such units. L2 units are collected from independent shadow-traffic groups without request-level pairing, whereas L3 and L4 units are collected from independently randomized online groups. The per-strategy mean and variance are
\begin{equation}
\bar{x}_{g,\ell}=\frac{1}{n_{g,\ell}}\sum_{i=1}^{n_{g,\ell}}x_{g,i}^{(\ell)},
\qquad
s_{g,\ell}^2=\frac{1}{n_{g,\ell}-1}\sum_{i=1}^{n_{g,\ell}}\bigl(x_{g,i}^{(\ell)}-\bar{x}_{g,\ell}\bigr)^2.
\end{equation}
The estimated effect and Welch standard error are
\begin{equation}
\widehat{\Delta}_{\ell}(c)=\bar{x}_{c,\ell}-\bar{x}_{b_{\ell},\ell},
\qquad
\mathrm{SE}_{\ell}(c)=\sqrt{\frac{s_{c,\ell}^2}{n_{c,\ell}}+\frac{s_{b_{\ell},\ell}^2}{n_{b_{\ell},\ell}}},
\end{equation}
with Welch--Satterthwaite degrees of freedom
\begin{equation}
\nu_{\ell}=\frac{\left(\dfrac{s_{c,\ell}^2}{n_{c,\ell}}+\dfrac{s_{b_{\ell},\ell}^2}{n_{b_{\ell},\ell}}\right)^2}
{\dfrac{\left(s_{c,\ell}^2/n_{c,\ell}\right)^2}{n_{c,\ell}-1}+\dfrac{\left(s_{b_{\ell},\ell}^2/n_{b_{\ell},\ell}\right)^2}{n_{b_{\ell},\ell}-1}}.
\end{equation}

\paragraph{Confidence interval.}
For every active level, the two-sided confidence interval at significance level $\alpha$ is
\begin{equation}
\mathrm{CI}_{\ell}(c)
=\left[
\widehat{\Delta}_{\ell}(c)-t_{\nu_{\ell},\,1-\alpha/2}\,\mathrm{SE}_{\ell}(c),\;
\widehat{\Delta}_{\ell}(c)+t_{\nu_{\ell},\,1-\alpha/2}\,\mathrm{SE}_{\ell}(c)
\right].
\end{equation}
When a relative effect is reported, the effect and interval endpoints are normalized by the corresponding baseline mean. By default, $\alpha=0.05$, corresponding to $95\%$ confidence intervals.

\paragraph{Promotion decision.}
With the evaluation signal oriented so that larger values indicate better outcomes, the promotion decision applies the interval endpoints $L_{\ell}(c)$ and $U_{\ell}(c)$ as in Section~\ref{sec:verifier-ladder}. A candidate is promoted when the interval lies entirely above zero ($L_{\ell}(c)>0$), which provides statistically significant evidence of a positive effect at the current significance level. It is retained for further evaluation when the interval contains zero ($L_{\ell}(c)\leq 0\leq U_{\ell}(c)$), which indicates that the observed effect is not yet distinguishable from no effect. It is stopped when the interval lies entirely below zero ($U_{\ell}(c)<0$), which provides statistically significant evidence of a negative effect. The relative interval $\mathrm{CI}_c^{\mathrm{rel}}$ shares the sign of the absolute interval and yields the same decision; it is reported for interpretability.

\section{Prompt Templates and Structured Outputs for Research-State Updates}
\label{app:core-prompts}

The following templates summarize the instructions for hypothesis revision and cross-round evidence reuse in \strategyharness. They retain the core inputs, update rules, and outputs while omitting deployment-specific identifiers and serialization details. 
\subsection{Hypothesis Update Prompt}
\label{app:hypothesis-prompt}

\begin{tcolorbox}[
    title=Prompt Template for Hypothesis Update,
    width=\textwidth,
    colframe=confblue,
    colback=confbluebg,
    colbacktitle=confblue,
    coltitle=white,
    fonttitle=\bfseries,
    boxrule=0.8pt,
    arc=1.5pt,
    breakable,
    fontupper=\small
]

\textbf{Objective and inputs.} Maintain testable parameter regularities across experimental rounds. Read the completed round summary, candidate records, confidence evidence, final review, lessons, and existing hypotheses.

\textbf{Update instructions.}
\begin{enumerate}[leftmargin=*,nosep]
    \item Generate at most three new hypotheses per round. Specify the parameter range, expected metric direction, risk boundary, and implications for candidate design; do not merely restate the best candidate.
    \item Merge equivalent claims with overlapping scopes. Record narrower findings as scoped evidence without broadening their support. Keep claims with different directions or risk conditions separate and link them through \texttt{Related}.
    \item Assign \texttt{support} when covered evidence supports the claim under the evaluation criteria, \texttt{refute} when it contradicts the claim or violates a decision constraint, and \texttt{neutral} when scope coverage or evidence is insufficient. Each round contributes at most one verdict per hypothesis.
    \item Increment the corresponding counter and set \texttt{Score = Support - Refute}. Scores of at least $2$ support \texttt{active} status; scores of at most $-2$ indicate \texttt{risky} hypotheses. Archive superseded or persistently unused hypotheses.
\end{enumerate}

\textbf{Plan use.} Retrieve relevant \texttt{active} hypotheses with scores of at least $2$ and \texttt{risky} hypotheses with scores of at most $-2$. Require a matching strategy version and explain each hypothesis's effect on candidate design. If none applies, mark the round as exploratory. These scores summarize hypothesis evidence; they are not verifier confidence estimates.

\tcbline
\textbf{Hypothesis Output Template.}
\begin{Verbatim}[fontsize=\small,breaklines=true,breakanywhere=true]
### H-{id}: {testable parameter regularity}
- **Status:** active / risky / archived
- **Strategy Version:** {strategy fingerprint}
- **Score:** {Support - Refute}
- **Support:** {supporting-round count}
- **Refute:** {refuting-round count}
- **Neutral:** {neutral-round count}
- **Scope:** {parameter ranges and experimental conditions}
- **Claim:** {expected metric direction under these conditions}
- **Plan Use:** {implications for subsequent candidate design}
- **Risk:** {trade-offs, applicability limits, or counterexamples}
- **Evidence:** {round:candidate verdict; ...}
- **Related:** {related hypothesis identifiers or none}
- **Updated At:** {timestamp}
\end{Verbatim}

\textbf{Plan-Stage Output Template.}
\begin{Verbatim}[fontsize=\small,breaklines=true,breakanywhere=true]
Used hypotheses: H-{id} score={value}: {design implication}
Candidate design: {candidate}: {modification; hypothesis tested}
\end{Verbatim}
\end{tcolorbox}

\subsection{Cross-Round Memory Update Prompt}
\label{app:memory-prompt}

\begin{tcolorbox}[
    title=Prompt Template for Cross-Round Memory Update,
    width=\textwidth,
    colframe=confblue,
    colback=confbluebg,
    colbacktitle=confblue,
    coltitle=white,
    fonttitle=\bfseries,
    boxrule=0.8pt,
    arc=1.5pt,
    breakable,
    fontupper=\small
]

\textbf{Objective and inputs.} Maintain two complementary records: structured lessons preserve round-level facts and reasoning; compressed memory summarizes cross-round findings and guides subsequent candidate design. Use all round summaries for the current strategy and the previous memory.

\textbf{Update instructions.}
\begin{enumerate}[leftmargin=*,nosep]
    \item After the final review, append one non-failure lesson per round with an \texttt{Outcome} matching the recorded decision. Record execution failures separately as \texttt{crash} lessons with their causes.
    \item Preserve the strategy version, experimental context, decision rationale, and evidence references. When summaries disagree with measured evidence or experimental records, correct the summaries.
    \item Before planning, retrieve relevant lessons with a matching strategy version. Use supported findings, avoid repeatedly unsuccessful directions, and resolve recurring execution failures. Incompatible lessons remain available for audit, not direct candidate guidance.
    \item Rewrite compressed memory after each round using the current strategy's summaries and preceding memory. Preserve current-run lessons, down-weight older findings, and compress related historical lessons without losing evidence references. Provide concrete guidance for subsequent candidate design.
\end{enumerate}

\textbf{Outcome semantics.} \texttt{keep}, \texttt{discard}, and \texttt{iterate} summarize the recorded round decision; \texttt{pivot} denotes a change of strategy family, \texttt{crash} an execution failure, and \texttt{summary} a historical synthesis.

\tcbline
\textbf{Lesson Output Template.}
\begin{Verbatim}[fontsize=\small,breaklines=true,breakanywhere=true]
### L-{N}: {title}
- **Strategy:** {candidate settings and search design}
- **Strategy Version:** {strategy fingerprint}
- **Outcome:** keep / discard / iterate / pivot / crash / summary
- **Insight:** {actionable finding to reuse or avoid}
- **Context:** goal=...; scope=...; metric=...; direction=...
- **Iteration:** {round identifier}#{task identifier}
- **Reason:** {rationale consistent with the recorded decision}
- **Refs:** trace={record}; experiment={id}; round={id}
- **Timestamp:** {timestamp}
\end{Verbatim}

\textbf{Compressed Memory.} The source protocol specifies its content rather than a fixed output schema: reusable findings, their applicability conditions and evidence references, and concrete candidate-design guidance for the next round.
\end{tcolorbox}

\subsection{Structured Experimental Records}
\label{app:cycle-output}

\begin{tcolorbox}[
    title=Structured Output Templates for Experimental Records,
    width=\textwidth,
    colframe=confblue,
    colback=confbluebg,
    colbacktitle=confblue,
    coltitle=white,
    fonttitle=\bfseries,
    boxrule=0.8pt,
    arc=1.5pt,
    breakable,
    fontupper=\small
]

\textbf{Recording instructions.} Link candidate designs, confidence evidence, and the final review through shared task and experimental-round identifiers. Retain one baseline per round and record each candidate's parameters and design rationale. Base selection and the final decision on the confidence-evaluation record, not legacy score logs. The templates below retain selected source fields; placeholders replace deployment-specific values.

\textbf{State and decision.} The source protocol uses \texttt{cycle} for an experimental round. Its state is \texttt{RUNNING}, \texttt{CLOSED}, or \texttt{BLOCKED}. Completed rounds record \texttt{KEEP}, \texttt{DISCARD}, or \texttt{ITERATE}, mapped to the corresponding lesson outcomes. Blocked rounds record a failure cause without a decision. These round-level labels are distinct from the verifier promotion labels in the Method.

\tcbline
\textbf{Final Review Template.}
\begin{Verbatim}[fontsize=\small,breaklines=true,breakanywhere=true]
{
  "task_name": "<task identifier>",
  "cycle_id": "<round identifier>",
  "stage": "review",
  "status": "DONE",
  "decision": "<KEEP | DISCARD | ITERATE>",
  "reason": "<evidence-based decision rationale>",
  "best_version": "<selected candidate identifier>",
  "best_selected_by": "llm_from_confidence_pipeline",
  "confidence_method": "<evaluation method>",
  "confidence_file": "<confidence-evidence record>",
  "best_confidence": {},
  "strategy_contract_version": "<strategy fingerprint>",
  "versions_file": "<candidate-design records>"
}
\end{Verbatim}

\textbf{Evidence Trace Template.}
\begin{Verbatim}[fontsize=\small,breaklines=true,breakanywhere=true]
{
  "task_name": "<task identifier>",
  "round": "<round label>",
  "cycle": {
    "cycle_id": "<round identifier>",
    "status": "CLOSED",
    "best_version": "<selected candidate identifier>",
    "confidence_method": "<evaluation method>",
    "confidence_file": "<confidence-evidence record>",
    "best_confidence": {},
    "decision": "<KEEP | DISCARD | ITERATE>",
    "versions_file": "<candidate-design records>"
  },
  "decision": "<same round decision>"
}
\end{Verbatim}
The \texttt{best\_confidence} object contains the selected candidate's confidence evidence. Candidate, decision, and evidence references must agree across the review, trace, and lesson records.
\end{tcolorbox}

\subsection{Knowledge-Augmented Candidate Generation Prompt}
\label{app:knowledge-augmented-prompt}

This template summarizes the Bayesian-optimization-inspired reasoning principles $\mathcal{K}_{\mathrm{BO}}$ used in Equation~\ref{eq:knowledge-augmented-proposal}. These principles guide qualitative reasoning about observed effects, uncertainty, and candidate diversity.

\begin{tcolorbox}[
    title=Prompt Template for Knowledge-Augmented Candidate Generation,
    width=\textwidth,
    colframe=confblue,
    colback=confbluebg,
    colbacktitle=confblue,
    coltitle=white,
    fonttitle=\bfseries,
    boxrule=0.8pt,
    arc=1.5pt,
    breakable,
    fontupper=\small
]
\textbf{Inputs.} Mutable parameters and their types, intervention bounds, baseline, candidate budget, historical evaluations, hypotheses, memory, and exact-replication requirements.

\textbf{Generation instructions.}
\begin{enumerate}[leftmargin=*,nosep]
    \item Classify the parameter space and diagnose the search state as \texttt{cold\_start}, \texttt{local\_refine}, \texttt{stagnant}, or \texttt{pivot}. Use repeated evidence, coverage, and contradictions rather than the largest point estimate alone.
    \item Select one reasoning strategy. Use \textbf{GP-inspired} reasoning for low-dimensional continuous spaces: contrast supported regions with uncertainty gaps. Use \textbf{TPE-inspired} reasoning for mixed continuous and integer spaces: distinguish supported, unfavorable, and unresolved parameter patterns. Use \textbf{SMAC-inspired} reasoning for categorical or conditional spaces: compare supported branches and structural counterfactuals.
    \item Form a candidate pool at least twice the available new-candidate budget. Cover refinement, boundary extension, uncertainty probes, and interaction tests. Preserve negative and inconclusive evidence when selecting candidates.
    \item Under stagnation or a change of search direction, increase exploration. Include a feasible point beyond the explored range on a meaningful axis and an interaction test or uncovered branch. Remain within the intervention bounds.
    \item Remove invalid, excluded, or duplicate proposals and select a diverse subset. Keep exact replications separate from new candidates: replication preserves the full configuration, while every new configuration is generated through the selected reasoning strategy.
\end{enumerate}

\tcbline
\textbf{Output Template.} Report concise, evidence-backed selection summaries and candidate roles. The following layout summarizes the required fields without reproducing the full planning schema.
\begin{Verbatim}[fontsize=\small,breaklines=true,breakanywhere=true]
ideation_strategy.method: <GP-/TPE-/SMAC-inspired>
search_state: <cold_start | local_refine | stagnant | pivot>
cycle_mode: <current experiment-allocation mode>
exploration_level: <level; high for stagnant or pivot>
selection_reason: <brief evidence-backed rationale>
Allocation: <exact-confirmation and new-candidate counts>
For each candidate:
  candidate_source: <confirmation_queue | algorithm>
  search_role: <replication or method-specific exploration role>
\end{Verbatim}
\end{tcolorbox}

\end{document}